\documentclass[a4paper]{article}
\def\kz {\ket{0}}
\def\ku {\ket{1}}

\def\kp#1{|\psi_{#1}\rangle}

\def\ket#1{| #1\rangle}               

\usepackage{graphicx}
\usepackage{epsfig,psfrag}
\usepackage[small]{caption2}
\usepackage[below]{placeins}
\usepackage{flafter}
\usepackage{amssymb}

\begin{document}

\title{Comments on quant-ph/0506137: Fast quantum search algorithms
by qubit comparisons exploiting global phase interference}
\author{L.A.B. Kowada$\dagger$\footnote{Corresponding author: kowada@cos.ufrj.br.}\ ,
C.M.H. de Figueiredo$\dagger$,
R. Portugal$\ddagger$,
and C.C. Lavor$^+$}
\date{$\dagger$ Programa de Engenharia de Sistemas e Computa\c
c\~ao/UFRJ\\%
$\ddagger$ LNCC/MCT\\%
$^+$ IMECC/UNICAMP}

\maketitle


 Recently, Andreas de Vries~\cite{vries} proposed a quantum algorithm
that would find an element in an unsorted database exponentially
faster  than Grover's algorithm~\cite{grover}. We show that de
Vries' algorithm does not work as intended and does not give any
clue about the position of the searched element.

\

The main part of the algorithm is depicted in figure~1, which is
reprinted from the original version of~\cite{vries}. The correct
output of the algorithm is
\begin{equation}
\kp{}=\frac{1}{\sqrt{2^n}}\sum_{k=0}^{2^n-1}(-1)^{f(k)}\ket{k}\ket{k},
\label{eq:output}
\end{equation}
where $f$ is the function that marks the searched elements as in
Grover's algorithm. The state~(\ref{eq:output}) yields no
information about $f$ upon a measurement in the computational
basis. The probability of getting the searched element before or
after the execution of the algorithm is the same.

\begin{figure}[htp]
  \psfrag{reg}{\begin{tabular}{c}Register of\\ $n$ qubits\end{tabular}}
  \psfrag{kz}{$\kz$}
  \psfrag{Hn}{$H^{\otimes n}$}
  \psfrag{P0}{$\stackrel{\uparrow}{\kp{0}}$}
  \psfrag{P1}{$\stackrel{\uparrow}{\kp{1}}$}
  \psfrag{P2}{$\stackrel{\uparrow}{\kp{2}}$}
  \psfrag{P2a}{$\stackrel{\uparrow}{\kp{2a}}$}
  \psfrag{P3}{$\stackrel{\uparrow}{\kp{3}}$}
  \psfrag{Qf}{$\begin{array}{c} Q_f\\ \ket{x}\rightarrow(-1)^{f(x)}\ket{x}\end{array}$}
 \centering \epsfig{figure=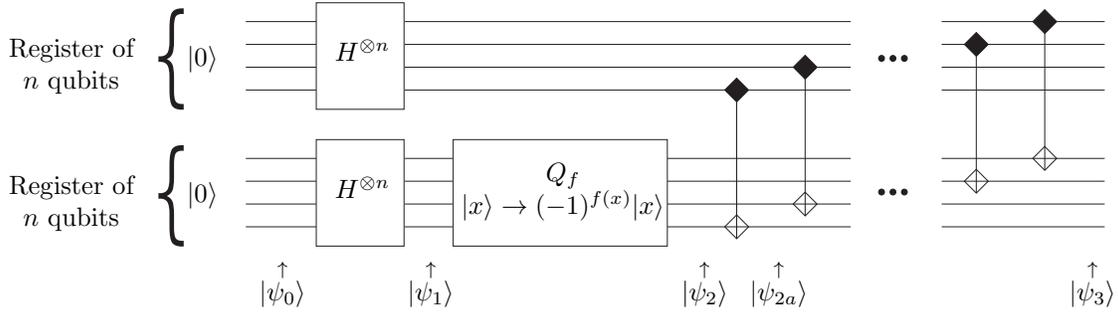,width=15cm}
  \caption{This circuit is taken from the original version of~\cite{vries}.}
\label{fig:vries5}
\end{figure}

We calculate below each state of the quantum computer as described
in the circuit of figure~\ref{fig:vries5}. For the states $\kp{0}$
and $\kp{1}$, we have
\begin{equation}
\kp{0} =\kz\kz
\end{equation}
and
\begin{equation}
\kp{1}=\frac{1}{2^n}\left(\sum_{j=0}^{2^n-1}\ket{j}\right)\left(\sum_{k=0}^{2^n-1}\ket{k}\right).
\end{equation}
The operator $Q_f$ acts, in fact, on $n+1$ qubits
\begin{equation}
Q_f\ket{k}\ket{-}=(-1)^{f(k)}\ket{k}\ket{-},
\end{equation}
but one can drop the last qubit for simplicity. See the details in
\cite{grover}. Then
\begin{equation}
\kp{2}=\frac{1}{2^n}\left(\sum_{j=0}^{2^n-1}\ket{j}\right)\left(\sum_{k=0}^{2^n-1}(-1)^{f(k)}\ket{k}\right).
\label{eq:psi2}
\end{equation}
Converting state $\kp{2}$ to binary notation we get
\begin{equation}
\kp{2}=\frac{1}{2^n}\sum_{j,k}(-1)^{f(k)}\ket{j_1}\cdots\ket{j_n}\ket{k_1}\cdots\ket{k_n},
\end{equation}
where $j=j_1\cdots j_n$ is the binary representation of $j$ and $\sum_j$ corresponds to $n$ sums running from 0 to 1. The same representation is used for $k$ and $\sum_k$.

The operator described by the filled diamond connected to a hollow diamond is given by
\begin{equation}
C_{i,i+n}=\frac{1}{\sqrt{2}}
\left(\begin{array}{cccc}
1 & 0 & 1 & 0\\
0 & 1 & 0 & -1\\
-1 & 0 & 1 & 0\\
0 & 1 & 0 & 1\\
\end{array}\right),
\end{equation}
where the subindices $i$ and $i+n$ describe in which qubits the operator is acting. Note that the action of $C_{i,i+n}$ on two qubits is given by
\begin{equation}
C_{i,i+n}\ket{j_i}\ket{k_i}=(-1)^{j_i k_i}\frac{\kz+(-1)^{1+j_i+k_i}\ku}{\sqrt{2}}\ket{k_i}.
\end{equation}
Using this equation, we can calculate the next state of figure 1.
Applying $C_{n,2n}$ to state~(\ref{eq:psi2}) we get
\begin{equation}
\kp{2a}=\frac{1}{2^n}\sum_{j,k}(-1)^{f(k)}(-1)^{j_n k_n}\ket{j_1}\cdots\ket{j_{n-1}}\frac{\kz+(-1)^{1+j_n+k_n}\ku}{\sqrt{2}}\ket{k_1}\cdots\ket{k_n}.
\end{equation}
After applying $C_{n-1,2n-1}$; ... ; $C_{1,n+1}$ we get the last
state of the circuit
\begin{equation}
\kp{3}=\frac{1}{2^n}\sum_{j,k}(-1)^{f(k)}(-1)^{j\cdot k}\frac{\kz+(-1)^{1+j_1+k_1}\ku}{\sqrt{2}}\cdots\frac{\kz+(-1)^{1+j_n+k_n}\ku}{\sqrt{2}}\ket{k_1}\cdots\ket{k_n},
\end{equation}
where $j\cdot k=j_1 k_1 + \cdots + j_n k_n$. Expanding the right hand side, we get
\begin{equation}
\kp{3}=\frac{1}{2^{n}\sqrt{2^{n}}}\sum_{j,k}(-1)^{f(k)}(-1)^{j\cdot k}\sum_l (-1)^{l_1+\cdots+l_n+l\cdot(j+k)}\ket{l_1}\cdots\ket{l_n}\ket{k_1}\cdots\ket{k_n}.
\end{equation}
Interchanging the order of the sums and using that
\begin{equation}
\sum_j(-1)^{j\cdot(k+l)}=2^n\delta_{kl},
\end{equation}
we get expression~(\ref{eq:output}).

\bibliographystyle{plain}

\end{document}